\documentclass[12pt,a4paper,twoside]{article}
\usepackage[round,sort]{natbib}
\usepackage[english]{babel}
\usepackage{url}
\usepackage[pdftex]{graphicx}
\usepackage{adjustbox}
\usepackage[svgnames,fixpdftex]{xcolor}
\usepackage{float}
\usepackage{fancyhdr}
\usepackage[Bjornstrup]{fncychap}
\usepackage{titling}
\usepackage{subcaption}
\usepackage{amsmath, amssymb, amsfonts}
\usepackage{bm}
\usepackage{mathtools}
\usepackage{clock}
\usepackage[T1]{fontenc}
\usepackage[applemac]{inputenc}
\usepackage{setspace}
\usepackage{xspace}
\usepackage[pdftex,left=2.5cm+12pt,right=2.5cm-10pt,top=2.5cm,bottom=2.5cm, headheight=24pt]{geometry}
\usepackage[labelfont=bf,labelsep=period,font=small,textfont=sl,width=0.85\textwidth, tableposition=top]{caption}
\usepackage[amsmath,thmmarks]{ntheorem}
\usepackage[explicit, outermarks]{titlesec}
\usepackage{enumitem}
\usepackage{tcolorbox}
\usepackage{booktabs}
\usepackage{rotating,tabularx}
\usepackage{multirow}
\usepackage{appendix}
\usepackage{verbatim}
\usepackage[babel,kerning, protrusion=allmath]{microtype}
\usepackage[scaled=0.95]{helvet,inconsolata}
\usepackage[autostyle, english=british]{csquotes} 

\definecolor{ao}{rgb}{0.0, 0.0, 1.0}

\newcommand{\helv}{\fontfamily{phv}\fontshape{n} \selectfont}
\newcommand{\ava}{\fontfamily{pag}\selectfont}

\titleformat{\section}{\ava \sffamily \bfseries \scshape \color{MidnightBlue}}{\thesection}{5pt}{#1}
\titleformat{\subsection}{\ava \sffamily \bfseries \color{MidnightBlue}}{\thesubsection}{5pt}{#1}
\titleformat{\subsubsection}{\ava\sffamily \color{MidnightBlue}}{\thesubsubsection}{5pt}{#1}
\titlespacing*{\chapter} {0pt}{20pt}{20pt}
\titlespacing*{\section} {0pt}{2.5ex plus 0.5ex minus .2ex}{2ex plus .2ex minus 0.3ex}
\titlespacing*{\subsection} {0pt}{2.25ex plus 0.5ex minus .2ex}{1.5ex plus .2ex minus 0.3ex}
\titlespacing*{\subsubsection}{0pt}{2.25ex plus 0.5ex minus .2ex}{1.5ex plus .2ex minus 0.3ex}

\DeclarePairedDelimiterX\pafun[2](){#1\nonscript\>\delimsize\vert\nonscript\> \mathopen{}#2}
\DeclarePairedDelimiterX\papro[2][]{#1\nonscript\>\delimsize\vert\nonscript\> \mathopen{}#2}
\DeclarePairedDelimiterX\paden[3](){#1\nonscript\>\delimsize\vert\nonscript\> \mathopen{}#2,#3}
\DeclarePairedDelimiterX\pastu[4](){#1\nonscript\>\delimsize\vert\nonscript\> \mathopen{}#2,#3,#4}
\DeclarePairedDelimiterX\palik[2](){#1\nonscript\>;\> \mathopen{}#2}
\DeclarePairedDelimiterX{\cb}[1]\{\}{#1}
\DeclarePairedDelimiterX{\rb}[1](){#1}
\DeclarePairedDelimiterX{\sqb}[1][]{#1}
\DeclarePairedDelimiter{\abs}{\lvert}{\rvert} 

\newcommand{\sqbra}[1]{\sqb*{#1}\xspace}
\newcommand{\paren}[1]{\rb*{#1}\xspace}
\newcommand{\cubra}[1]{\cb*{#1}\xspace}
\newcommand{\set}[1]{\ensuremath{\cb*{#1}}\xspace}

\DeclareMathOperator*{\argmin}{arg\,min}  
 \DeclareMathOperator{\var}{\mathbb V} \DeclareMathOperator{\E}{\mathbb E}    \DeclareMathOperator{\los}{\mathcal L} 
 \DeclareMathOperator{\p}{P} \DeclareMathOperator{\ver}{L}

\DeclareMathOperator{\vr}{VaR}

\newcommand{\Real}{\ensuremath{\mathbb R\xspace}}
\newcommand{\bl}[1]{\ensuremath{\boldsymbol{#1}}\xspace}
\newcommand{\samp}[3][1]{\ensuremath{\cubra{#3_{#1},\dots,#3_{#2}}}\xspace}

\newcommand{\gorro}[1]{\ensuremath{\hat{#1}}\xspace}

\newcommand{\dd}[1]{\ensuremath{~\mathrm{d}{#1}}}

\newcommand\mo[3][f]{\ensuremath{#1\pafun*{#2}{#3}}\xspace}
\newcommand{\pri}[1][\theta]{\ensuremath{\pi\paren{#1}}\xspace}
\newcommand{\mop}[3][\pi]{\mo[#1]{#2}{#3}\xspace}
\newcommand{\cp}[2]{\ensuremath{\p\papro*{#1}{#2}}\xspace}
\newcommand{\ce}[2]{\ensuremath{\E\papro*{#1}{#2}}\xspace}
\newcommand{\cv}[2]{\ensuremath{\var\papro*{#1}{#2}}\xspace}
\newcommand{\lik}[2]{\ensuremath{\ver\palik*{#1}{#2}}\xspace}

\newcommand{\ga}[3][x]{\ensuremath{\mathrm{Ga} \paden*{#1}{#2}{#3}}\xspace}
\newcommand{\ig}[3][x]{\ensuremath{\mathrm{IGa}\paden*{#1}{#2}{#3}}\xspace}

\newcommand{\be}[3][x]{\ensuremath{\mathrm{Be} \paden*{#1}{#2}{#3}}\xspace}
\newcommand{\ip}[3][x]{\ensuremath{\mathrm{iPa} \paden*{#1}{#2}{#3}}\xspace}
\newcommand{\pa}[3][x]{\ensuremath{\mathrm{Pa} \paden*{#1}{#2}{#3}}\xspace}
\newcommand{\ex}[2][x]{\ensuremath{\mathrm{Ex} \pafun*{#1}{#2}}\xspace}

\newcommand{\un}[3][x]{\ensuremath{\mathrm{Un} \paden*{#1}{#2}{#3}}\xspace}

\newcommand{\lex}[3][x]{\ensuremath{\mathrm{lEx} \paden*{#1}{#2}{#3}}\xspace}

\newcommand{\gpd}[3][x]{\ensuremath{\mathrm{gPa} \paden*{#1}{#2}{#3}}\xspace}
\newcommand{\paga}[2]{\ensuremath{\mathrm{PG}\pafun*{#1}{#2}}\xspace}

\newcommand{\refi}[1]{Figure~\ref{fig:#1}\xspace}
\newcommand{\reta}[1]{Table~\ref{tab:#1}\xspace}
\newcommand{\rese}[1]{Section~\ref{sec:#1}\xspace}
\newcommand{\resub}[1]{Section~\ref{sub:#1}\xspace}

\newcommand{\ie}{\emph{i.e.}\xspace}
\newcommand{\eg}{\emph{e.g.}\xspace}

\newcommand{\upd}[1]{\ensuremath{{#1}^{\star}}\xspace}

\graphicspath{{Figures/}}
\theorembodyfont{\helv} \theoremstyle{plain}
\newtheorem{definition}{Definition}

\allowdisplaybreaks
\tcbset{colframe=blue!75!black, colback=blue!5!white, fonttitle=\large\ava\bfseries, center title, fontupper=\normalsize\ava, width=0.9\textwidth}
\setlist[itemize]{itemsep=3pt}
\setlist[enumerate]{itemsep=3pt}
\setlist[description]{itemsep=3pt}

\author{J. Sharpe \\ Sharpe Actuarial ltd \and M. A. Juárez \\ University of Sheffield}

\title{Calibration of the Pareto and related distributions---a reference-intrinsic approach}

\date{}

\begin{document}
\maketitle
\vspace*{-2em}
\begin{abstract}
	We study two Bayesian (Reference Intrinsic and Jeffreys prior) and two frequentist (MLE and PWM) approaches to calibrating the  Pareto and related distributions.  Three of these approaches are compared in a simulation study and all four to investigate how much equity risk capital banks subject to Basel II banking regulations must hold.  	The Reference Intrinsic approach, which is invariant under one-to-one transformations of the data and parameter, performs better when fitting a generalised Pareto distribution to data simulated from a Pareto distribution and is competitive in the case study on equity capital requirements.  
	\\[2ex]
	
	\noindent \textbf{Keyowrds} Generalised Pareto distribution, Intrinsic loss, invariance, Reference prior. \\[2ex]
	\noindent \textbf{Correspondence details} \\[1ex]
	J. Sharpe, Sharpe Actuarial ltd.  London, UK.  \texttt{james@sharpeactuarial.co.uk} \\
	Miguel A. Juárez, School of Mathematics and Statistics, University of Sheffield, S3 7RH, Sheffield, UK. \texttt{m.juarez@sheffield.ac.uk}
	
\end{abstract}

\onehalfspace

\section{The Generalised Pareto and related distributions} \label{sec:related}

The Generalised Pareto distribution (GPD) is widely used in engineering, environmental science and finance to model low probability events.  Typically the GPD is used to estimate extreme percentiles such as the 99th percentile for a specific event.  This might be used for setting the height of flood wall defences or estimating how much capital banks might hold for specific market risks.

We say that the positive quantity $x$ follows a GPD if it has probability density function (PDF)
\begin{equation}
	\gpd\kappa\sigma = \frac{1}{\sigma}  \begin{cases}\paren{1 - \kappa/\sigma\, x}^{1/\kappa-1} & \kappa \not = 0 \\ \exp\paren{-x/\sigma} & \kappa=0\end{cases}, \label{gpd}
\end{equation}
with $\sigma>0$.  The support of the distribution $\mathcal X = (0, \infty)$ if $\kappa \le 0$, while $\mathcal X = (0, \sigma/\kappa]$ if $\kappa >0$; thus $\kappa$ is a shape and $\sigma$ is scale parameter.  The mean, 
\[
	\ce{x}{\kappa, \sigma} = \frac\sigma{1 + \kappa} \quad \text{and variance} \quad \cv x{\kappa, \sigma} = \frac{\sigma^2}{(\kappa+1)^2 (2 \kappa +1)},
\]
exist iff $\kappa> -1$ and $\kappa >-1/2$, respectively.

The GPD is related to several distributions.  It clearly has an exponential distribution with mean $\sigma$ as a special case when $\kappa =0$.  Further, $y = \sigma - \kappa x$ follows a Pareto distribution, \pa[y]{-1/\kappa}{1/\sigma}, if $\kappa<0$, where
\begin{equation}
	\pa[y]{\alpha}{\beta} = \frac\alpha\beta \paren{\frac y\beta}^{-(\alpha + 1)}; \qquad y \ge \beta, \quad \alpha, \beta >0, \label{padi}
\end{equation} 
and an inverted-Pareto, \ip[y]{1/\kappa}{\sigma}, if $\kappa>0$, where
\begin{equation}
	\ip[y]\alpha\beta = \frac\alpha\beta \paren{\frac y\beta}^{\alpha - 1}~; \qquad y \le \beta, \quad \alpha, \beta >0. \label{ipdi}
\end{equation}
In the latter case, $z=y^{\alpha}$, distributes uniformly on $\paren{0,\beta^\alpha}$; $z = y/\beta$ follows a Beta distribution, \be[z]{\alpha}{1}; and  $z = -\log y$ follows a location (or shifted) Exponential distribution,
\[
	\lex[z]{\alpha}{\theta} = \alpha \, \exp \sqbra{- \alpha \paren{z - \theta}}; \qquad z \ge \theta, \quad \alpha >0 , \theta \in \Real,
\]
where $\theta =\log \beta$. If the shape is fixed, $z = \log(\beta/y)$ follows an Exponential distribution with rate $\alpha$, \ex[z]{\alpha}. Similarly, if $y$ follows a \pa[y]{\kappa}{\sigma}, then $x = \kappa \paren{\sigma - y}$ follows a \gpd{-1/\kappa}{\sigma}. If a sample, \bl y = \samp ny, from the Pareto distribution is available, $\set{y_{(1)}, t}$ is sufficient, with $y_{(1)} = \min\set{\bl y}$ and $t = \prod_{i=1}^n y_i$.  No such sufficient statistics exist for the GPD.

One key feature of this family of distributions is their so-called lack of memory, a property at the core of their prominence in extreme value theory, related to peaks-over-threshold theory described in \resub{backg}.  Specifically let $x \thicksim \gpd\kappa\sigma$ and consider \cp{x \ge t+u}{x \ge t}, it is straightforward to prove that $u \thicksim \gpd[u]\kappa{\sigma'}$, with $\sigma' = \sigma-\kappa t$, hence $\ce{x-u}{x>t} = \sigma'/(1 + \kappa)$,  which is commonly used to graphically check model fit \citep{davidson}. It is immediate to check that if $x$ follows a \pa{\alpha}{\beta} then $\ce{x-u}{x>t} = t/(\alpha-1)$, provided $\alpha>1$, and thus a similar graphical model fit check can be carried out. 

In this paper we consider the Bayesian Reference Intrinsic (BRI) approach \citep{io, rueda, intint} for calibrating this family of distributions, and compare it with three alternative approaches, Maximum Likelihood (ML), Probability Weighted Moments (PWM) and a Bayesian approach using a Jeffreys prior, implemented using Markov chain Monte Carlo (MCMC). A simulation study is carried out comparing the average mean square error of three of the four approaches when calibrated to synthetic data from a Pareto distribution.  These methods are then applied to some equity return data in a case study and the results compared.

\subsection{The GPD and extremes} \label{sub:backg}

The GPD was first introduced by \citet{Pickands} in the extreme value framework as a distribution of sample excesses over a sufficiently high threshold \citep{zea1, zea2}.  Two key theories of the extreme value framework the GPD arose from are summarised below ---we use the second theory in the case study in \rese{real}.

\subsubsection{Extreme Value Theory 1}
Let $X_n$ be a sequence of iid random variables.  If these are divided into blocks of size $k$, $\bl x_j = \samp[(j-1)k+1]{jk}{x}$ and $M_j = \max\set{ \bl x_j}, ~ j=1, \dots, n/k$ (\ie the largest value in each block), then the $M_j$ follow a Generalised Extreme Value (GEV) distribution, with cumulative distribution function (CDF)
\[
	\mo[F]{m}{\kappa, \sigma} = \begin{cases}\exp\sqbra{- \paren{1 - \frac{km}\sigma}^{1/k}} & \kappa \not= 0 \\ \exp\sqbra{\exp(-m)} & \kappa =0\end{cases},
\]
with $\sigma>0$ the scale parameter and $\kappa$ the shape parameter \citep[][p.~265]{frey}.  Thus, the distribution of the maxima of blocks of data from almost any probability distribution follows a GEV with some shape parameter $\kappa$.

\subsubsection{Extreme Value Theory 2 (Picklands-Balkema-de Haan)} \label{ssub:pickands}
Let $X>0$ be a random quantity with CDF $F$.  The excess over threshold $u$ has CDF
\[
	F_u(x) = \cp{X-u \le x}{x>u} = \frac{F(x +u) - F(u)}{1 - F(u)},
\]
for $0 \le x \le x_F - u$,  where $x_F>0$ is the upper bound of the support of $F$.

There is a positive measurable function $B(u)$ such that
\[
	\lim_{u \to x_F} \sup_{x \le x_F - u} \abs*{F_u(x) - \gpd \kappa{B(u)}} =0,
\]
where $\kappa$ is the shape parameter of the GPD and $B(u)$ is the scale parameter, which is a function of the threshold  \citep[][p.~277]{frey}.  This means that whilst the scale parameter changes as the threshold changes, the shape parameter stays the same.  The distributions for which the block maxima converge to a GEV distribution constitute a set of distributions for which the excess distribution converges to the GPD as the threshold is raised.  The shape parameter of the GPD of the excesses is the same as the shape parameter of the GEV of block maxima.  This means that the excess above a threshold can effectively be modelled by a GPD (almost) regardless of the distribution of the full data set as long as the threshold is high enough.  This feature is used in a case study in \resub{peaks}, where the GPD is calibrated to just the tail of the data.

Characterising the GPD and deriving probabilistic and statistical results are extensively addressed in the literature \citep[\eg see][and references therein]{gala2,leadbetter, beirlant, Embrechts, Coles, Kotz, Castillo, Haan}.  Several approaches have been proposed to calibrate the GPD mainly focussing on the MLE, PWM or Method of Moments (MoM) \citep[see \eg][and references therein]{zea1, zea2}.  More recently, Bayesian approaches have been investigated \citep[see \eg][]{Lima, Ragulina, Juarez, Tancredi, Turkman}. \citet{Gilleland} reviews available software for estimation.  We refer the reader to \citet{zea1, zea2} which include summary tables of papers describing how the GPD is calibrated to a wide range of data sets, reproduced in \reta{paps} in the Appendix to show some of the extensive literature covering calibration of the GPD.

\subsection{The Pareto principle: the Lorenz curve and Gini index} \label{sec:gini}

The \enquote{80-20 rule} or Pareto principle has reached popular culture through books such as \citet{Koch}.  It is a way of more easily explaining the calibration of a Pareto or GPD.  An example of this is the 80-20 rule identified by V. Pareto in 1897 that 20\% of the population had 80\% of the wealth \citep{Persky}.  It is possible to use the calibration of the GPD to identify the Pareto principle parameters through the Lorenz curve and Gini index.  The Lorenz curve,
\[
	L(u) = \frac{1}{\mu} \, \int_0^u F^{-1}(z) \dd z, \qquad u \in (0,1),
\]
where $F(x)$ is the CDF of the random quantity $x$  and $\mu$ its expected value, describes precisely this relationship.  In case the distribution of the size is homogenous, \ie \enquote{$u$\% of the population accumulates $u$\% of the income}, then $L(u) =u$.  This motivates some measures of inequality, such as the Gini index, 
\[
	G = 1 - 2 \int_0^1 L(u) \dd u,
\]
which is the relative area under $L(u)$, respect to the straight line: the closer $G$ to 0(1), the more(less) egalitarian the distribution.  For the GPD,
\[
	L(u) = \frac 1\kappa \paren{(1-u)^{\kappa+1} + (\kappa+1) u -1} \quad \text{and} \quad G = (\kappa +2)^{-1},
\]
depend only on the shape parameter, thus inference on $L$ and/or $G$ is tantamount to inference on this parameter, which is explored in \rese{real}.

\section{Intrinsic calibration} \label{sec:bayes}

The Bayesian reference-intrinsic (BRI) approach \citep{io, rueda} provides a non-subjective Bayesian alternative to point estimation, based on the reference prior \citep{sun, sun2} and an intrinsic loss function \citep{Robert}.  In short,  Let \set{\mo{\bl x} \theta, \bl x \in \mathcal X, \theta \in \Theta} be a probability model assumed to describe the probabilistic behaviour of the observables $\bl x$, and suppose that a point estimator, $\theta^e = \theta(\bl x)$, of the parameter $\theta$ is required.  From a Bayesian decision standpoint, the optimal estimate, $\upd\theta$, minimises the expected loss,
\[
	\upd\theta = \argmin_{\theta^e \in \Theta } \int_\Theta \los\paren{\theta^e, \theta} \> \mop\theta{\bl x} \dd\theta,
\]
where $\los\paren{\theta^e, \theta}$ is a loss function measuring the consequences of estimating $\theta$ by $\theta^e$ and \mop\theta{\bl x} is the decision maker posterior distribution.  The BRI approach argues that in fact one is interested in using \mo{\bl x}{\theta^e} as a proxy of \mo{\bl x} \theta and thus the loss function should reflect this.  It advocates the use of the  Kullback–Leibler (KL) divergence as an appropriate measure of discrepancy between two distributions.  The KL (or directed logarithmic) divergence, 
\[
	\mo[K]{\theta_2}{\theta_1} = \int_{\mathcal X} \mo{\bl x}{\theta_1} \> \frac{\mo{\bl x}{\theta_1}}{ \mo{\bl x}{\theta_2}} \dd{\bl x},
\]
is nonnegative and nought if and only if $\theta_1 = \theta_2$, and it is invariant under one-to-one transformations of either \bl x or $\theta$.  However, the KL divergence is not symmetric and it diverges if the support of \mo{\bl x}{\theta_2} is a strict subset of the support of \mo{\bl x}{\theta_1}. To simultaneously address these two unwelcome features \citet{io} propose to use the intrinsic discrepancy,
\begin{equation}
	\delta(\theta, \theta^e) = \min\set{\mo[K]{\theta}{\theta^e}, \mo[K]{\theta^e}{\theta}}, \label{indis}
\end{equation}
a symmetrised version of the KL divergence.  This is taken as the quantity of interest, for which a reference posterior is derived.  The intrinsic estimator can then be obtained.

\begin{definition}[Bayesian intrinsic estimator] \label{intest}
	Let \set{\mo{\bl x}{\theta}, \bl x \in \mathcal X, \theta \in \Theta} be a family of probability models for some observable data \bl x, where the sample space, $\mathcal X$ may possibly depend on the parameter value. The BRI estimator,
\begin{gather*}
	\upd\theta(\bl x) = \argmin_{\theta^e \in\Theta} d(\theta^e, \bl x) 
	\shortintertext{where}
	d(\theta^e, \bl x)= \int_\Theta  \delta(\theta, \theta^e) \> \mop[\pi_\delta]\theta{\bl x} \dd\theta,
\end{gather*}
is the intrinsic expected loss and $\mop[\pi_\delta]\theta{\bl x}$ is the reference posterior for the intrinsic discrepancy, $\delta(\theta, \theta^e)$, as defined in~\eqref{indis}.
\end{definition}

Within the same methodology one can also obtain interval estimates, \ie credible regions.  
\begin{definition}[Bayesian intrinsic interval]\label{def:intint}
	Let \set{\mo{\bl x}{\theta}, \bl x \in \mathcal X, \theta \in \Theta} and $d(\theta^e, \bl x)$ be as in Definition~\ref{intest}.  A BRI interval, $R_p = R_p(\bl x) \subset \Theta$, of probability $p \in (0,1)$, is a subset of the parameter space $\Theta$ such that
	\[
		\int_{R_p} \mop[\pi_\delta]\theta{\bl x} \dd\theta =p \quad \text{and} \quad d(\theta_r, \bl x) \le d(\theta_s, \bl x),
	\]
	for all $\theta_r \in R_p$ and $\theta_s \not\in R_p$.
\end{definition}

BRI credible regions are typically unique and, since they are based in the invariant intrinsic discrepancy loss, they are also invariant under one-to-one transformations \citep{intint}.  

\subsection{Calibration for the Pareto family of  distributions} \label{sec:calib}

In Section~\ref{sub:backg} we highlighted the relationship between the GPD and the Pareto and Inverse Pareto distributions.  The main characteristic we will exploit here is that the GPD shape parameter remains invariant to any of those transformations, whilst the GPD scale parameter is linearly transformed. Given that the support of the inverted Pareto is bounded, it is easier to calibrate directly than the Pareto or the GPD.  For this reason we choose to work with this parameterisation and apply the results to the GPD parameters by the above simple transformations. 

Let $\bl x= \samp nx$ be a random sample from an \ip[x_i]{\kappa}{\sigma}, using \eqref{ipdi}, the likelihood is
\[
   \lik{\kappa,\sigma}{\bl x} \propto \kappa^n \sigma^{- n \kappa} t_1^{n \kappa} \,, \quad \sigma \ge t_2 ; \quad \text{where} \quad  \set{t_1, t_2} = \set{\prod_1^n x_i^{1/n}, x_{(n)}}
\]
are jointly sufficient, with $x_{(n)} = \max \samp nx$.  Moreover,  given \set{\kappa,\sigma} the MLE,
\begin{equation*}
	\gorro \sigma = t_2 \qquad \text{and} \qquad \gorro \kappa = \paren{\log\frac{t_2}{t_1}}^{-1}, \label{ipmle}
\end{equation*}
are conditionally independent, with sampling distributions \ip[\gorro \sigma]{n\kappa}{\sigma} and \ig[\gorro\kappa]{n}{n\kappa}, where the latter is an inverted Gamma distribution \citep{malik}.

The conjugate prior is a Pareto-Gamma distribution,
\begin{align}
	\paga{\kappa,\sigma}{k,b,c,d} &= \pa[\sigma]{k \kappa}{b} \: \ga[\kappa]{c}{d}, \quad \kappa>0, \sigma \ge b; \> k,b,c,d>0  \notag\\
	 &= k \kappa b^{k \kappa} \sigma^{-(k \kappa+1)} \frac{d^c}{\Gamma\sqbra{c}} \kappa^{c-1} \exp \sqbra{- d \kappa}.\label{paga}
\end{align}
which yields a Gamma marginal posterior \ga[\kappa]{n+c}{d+q}, where $q = n \gorro \kappa + k \log s$ and $ s = \max\set{\gorro\sigma, b}$.  For any choice of prior parameters, the posterior is asymptotically Gaussian and will converge to a mass point at \gorro \kappa a.s. as $n \to \infty$.

From \eqref{indis}, the intrinsic discrepancy for the inverted Pareto distribution, \ip\kappa\sigma, when the parameter of interest is the shape, can be written as
\begin{equation}
	\delta \paren{\kappa, \kappa^e} = n \begin{cases}- \log \theta + \theta -1 & \theta <1 \\ \log \theta + \theta^{-1} -1 & \theta \ge 1\end{cases}, \label{aldisc}
\end{equation}
with $\theta = \kappa/\kappa^e$, which does not depend on $\sigma$.  Following \citet{Juarez}, given that \eqref{aldisc} is a (piecewise) one-to-one function of $\kappa$, we can use the reference prior $\pri[\kappa,\sigma] \propto~\paren{\kappa \sigma}^{-1}$, a liming case of \eqref{paga}, which yields the marginal posterior \ga[\kappa]{n-1}{n/\gorro\kappa}, for $n>1$.   The intrinsic expected loss,
\[
	\mo[d]{\kappa_0}{\gorro \kappa} = \int_0^\infty \delta(\kappa,\kappa_0) \:  \ga[\kappa]{n-1}{n/\gorro\kappa} \dd \kappa,
\]
is defined for all $n>2$ and can be calculated numerically. Due to the asymptotic Gaussianity of the posterior, the approximations $\mo[d]{\kappa^e}{\gorro \kappa} \approx \delta\paren{\gorro\kappa,\kappa^e} + 1/2$ and $\upd \kappa\approx \gorro \kappa \paren{1-3/2n}$, work well even for moderate sample sizes.

The intrinsic discrepancy when the scale is the parameter of interest is
\[
	\delta\paren{\sigma,\kappa; \sigma^e} = n \; \begin{cases}\log(1 - \phi) & \phi <0 \\ \phi & \phi \ge 0\end{cases},
\]
where $\phi = \kappa \log \paren{\sigma/\sigma^e}$. The reference prior is $\pri[\phi, \kappa] \propto \kappa^{-1}$, or in terms of the original parameterisation, $\pri[\sigma,\kappa] \propto \sigma^{-1}$; which is not a limiting case of the Pareto-Gamma family.

In this case the loss function depends on both parameters and thus
\begin{gather}
	\bar \sigma = \argmin_{\sigma^e \ge \gorro \sigma} \mo[d]{\sigma^e}{\bl x} \label{iebeta}
	\shortintertext{with}
	\mo[d]{\sigma^e}{\bl x} = \int_{\gorro\sigma}^\infty \int_0^\infty  \delta\paren{\sigma,\kappa; \sigma^e}\;  \mop{\sigma, \kappa}{\bl x} \dd \kappa \dd \sigma	\label{isbeta}
	\shortintertext{and}
	\mop{\sigma, \kappa}{\bl x} = \frac{n^{n+1}}{{\gorro\kappa}^n \Gamma[n] } \, \kappa^{n}  \sigma^{-(n\kappa+1)} t_1^{n\kappa} \qquad \kappa>0, \, \sigma \ge \gorro\sigma	\notag
\end{gather}

The corresponding Bayes rule can be calculated numerically.  An analytical approximation, which works well even for moderate sample sizes, can be obtained by substituting the shape parameter with a consistent estimator in \eqref{isbeta}, carrying out the one dimensional integration and then solving \eqref{iebeta}.  Using the MLE, $\gorro\kappa$, yields $\bar \sigma \approx 2^{\frac{1}{n \gorro \kappa}}  \gorro\sigma$.

The Uniform and location-Exponential models are particular cases of the inverted Pareto (see \rese{related}). For the former, we have \un{0}{\sigma} = \ip{1}{\sigma} and in this case the intrinsic discrepancy is
\[
	\delta\paren{\sigma, \sigma^e} = n \; \abs*{\log \frac{\sigma}{\sigma^e}}
\]
and the corresponding reference prior is $\pri[\sigma] \propto \sigma^{-1}$, which yields a \pa[\sigma]{n}{\gorro \sigma} posterior, with $\gorro\sigma = x_{(n)}$ the MLE.  The expected intrinsic discrepancy has a simple analytical form,
\begin{align*}
	\mo[d]{\sigma^e}{\gorro \sigma} &= n \sqbra{\int_{\gorro\sigma}^{\sigma^e} \log \paren{\frac{\sigma^e}{\sigma}} n {\gorro \sigma}^n \sigma^{-(n+1)} \dd \sigma + \int^{\infty}_{\sigma^e} \log \paren{\frac{\sigma}{\sigma^e}} n {\gorro \sigma}^n \sigma^{-(n+1)} \dd \sigma} \\
	&= 2 z - \log z -1~,
\end{align*} 
where $z = \paren{\gorro\sigma/\sigma^e}^n$.  It is immediate to prove that the BRI estimator, $\tilde \sigma = 2^{\frac{1}{n}} \gorro\sigma$, is the median of the posterior, highlighting its invariance under one-to-one transformations. Indeed, the BRI estimator of the parameter in the location-Exponential model, \lex[y]{1}{\phi}, where $\phi = \log \sigma$, \ie the distribution obtained by letting $y = \log x$, is $\tilde\phi = \gorro\phi - n^{-1} \log 2 = \log\tilde\sigma$.

\section{Alternative approaches} \label{sec:other}
We briefly describe two alternative frequentist approaches, maximum likelihood estimate (MLE) and probability weighted moments (PWM).  We also describe a Bayesian approach that uses a Jeffreys prior which yields a proper posterior for any sample size.  The posterior has no analytical form, so we implement an MCMC strategy to sample from it.

\subsection{Jeffreys prior} \label{sub:mcmc}
For an alternative Bayesian approach, we use the independent Jeffreys prior,
\[
	\pri[\kappa, \sigma] \propto \sigma^{-1} \paren{1 - \kappa}^{-1} \paren{1 - 2\kappa}^{-1/2}, \quad \kappa < \frac 12, \sigma>0,
\]
which, despite being improper, yields a proper posterior for any sample size \citep{castellanos}.  The posterior, $\mop{\kappa, \sigma}{\bl x}$, is not analytical, so we implement an MCMC scheme to carry out inference.  Our strategy is a Metropolis-within-Gibbs algorithm with full conditionals
\begin{align*}
	\mop{\kappa}{\sigma, \bl x} &\propto \paren{1 - \kappa}^{-1} \paren{1 - 2\kappa}^{-1/2} \prod_{i=1}^n \paren{ 1 - \frac \kappa\sigma x_i}^{1/\kappa -1} \\
	\mop{\sigma}{\kappa, \bl x} &\propto \sigma^{-n +1} \prod_{i=1}^n \paren{ 1 - \frac \kappa\sigma x_i}^{1/\kappa -1}.
\end{align*}
To set the proposal distributions, one must bear in mind the parameter space depends on the sample space; specifically, $0 < x_i < \sigma/\kappa$ if $\kappa >0$.  Hence, for the shape parameter, $\kappa$, we propose from a truncated Gaussian with mode at the MLE, and upper bound at $\min \set{1/2, \sigma^c/x_{(n)}}$, where $\sigma^c$ is the current state of the shape; we use the free parameter to control the acceptance rate.  For the scale, $\sigma$, if the current state of the shape, $\kappa^c <0$, we use a Gamma proposal with mode at the $\sigma^c$ and use the free parameter to control the acceptance rate; otherwise, we propose from a truncated Gaussian with lower bound at $\kappa^c x_{(n)}$, mode at $\sigma^c$, and use the free parameter to control the acceptance rate.  Our \texttt R code is available under request from the corresponding author.

\subsection{Two frequentist approaches} \label{sub:twofreq}

\subsubsection{Maximum likelihood}\label{ssub:mle}
For a sample \bl x = \samp n x from a \gpd\kappa\sigma, the log-likelihood can be expressed as
\[
	\ell(\kappa, \sigma \>; \bl x) = \begin{cases}- n \log\sigma + \paren{\frac 1\kappa -1 } \sum_{i=1}^n \log \paren{1 - \frac \kappa\sigma x_i} & \kappa \not= 0 \\ - n \log \sigma + \frac n\sigma \bar x & \kappa = 0\end{cases}.
\]
The MLE exist only for $\kappa \le 1$ and is typically found using numerical methods.  If $\kappa < 1/2$, its sampling distribution is asymptotically Gaussian with covariance matrix \citep{zea1}
\[
	\frac 1n \paren{\begin{matrix} (1-\kappa)^2 & \sigma (1 - \kappa) \\ \sigma (1 - \kappa) & 2 \sigma^2(1- \kappa) \end{matrix}}.
\]

\subsubsection{Probability weighted moments}\label{ssub:pwm}

Probability weighted moments (PWM), or L-moments,
\[
	\mathrm M_{p,r,s} = \E \sqbra{ X^p F^r(x) \paren{1 - F(x)}^s}, \quad p,r,s \in \Real,
\]
characterise the distribution function $F$ of a random quantity $X$ and are exploited as a robust alternative to the method of moments for point estimation \citet{Greenwood}.  Particularly for the GPD, \citet{Diebolt}, suggest using 
\[
	\mu_s = \mathrm M_{1,0,s} = \E\sqbra{X \paren{1 - F(x)}^2} = \frac \sigma{(s+1) ( s+1 + \kappa)}, \quad \kappa>-1; \> s=0, 1,\dots
\]
from which 
\[
	\kappa =\frac{\mu_0}{\mu_0 - 2 \mu_1} -2 \quad \text{and} \quad \sigma = \frac{2 \mu_0 \mu_1}{\mu_0 - 2\mu_1}.
\]
The corresponding PWM estimators are obtained by substituting $\mu_0$ and $\mu_1$ by the estimators $\mu_j = n^{-1} \sum_{i=1}^n x_{(i)} \paren{1 - p_{(i)}}^s$, with $x_{(j)}$ the $j$-th order statistic.  Various expressions are available for $p_{(j)}$, in the sequel we use $p_{(j)} = (j+\gamma)/(n+c)$, with $\gamma=-0.35$ and $c=0$ as in \citet{zea1}.   For large sample sizes and if $-1 < \kappa < 1/2$, the PWM estimators are asymptotically Gaussian \citep{Diebolt} with covariance matrix
\[
	\frac{n^{-1}}{(1 + 2\kappa)(3+2\kappa)} \left(\begin{matrix} (1+\kappa) (2+\kappa)^2 (1 +\kappa +2\kappa^2) & \sigma (2+\kappa) ( 2+6\kappa +7\kappa^2 +2\kappa^3) \\
 \sigma (2+\kappa) ( 2+6\kappa +7\kappa^2 +2\kappa^3) & \sigma^2 (7 + 18\kappa +11\kappa^2 + 2\kappa^3)  \end{matrix}\right).
\]

\citet{land} have a different angle, exploiting the lack of memory to find a threshold and then estimate the shape parameter for the tail distribution of the size of agricultural land by county in the USA.

\section{Synthetic data and comparison}\label{sec:syn}

We carry out a simulation study to compare the calibration efficiency of the BRI, the ML and PWM estimators and present results on the shape parameter only for brevity.  As the MHA is itself a time consuming simulation process it has been left out of this comparison. We generate 10,000 samples of size $n=15, 50, 100$, from a Pareto distribution, \pa[x]\kappa\sigma ---which is linked to the GPD as described in \rese{related}---, with $\sigma=4$ and $\kappa = 1/3, 3, 7$.  The parameters are calibrated using the BRI, ML and PWM approaches, their sampling distribution are estimated and their bias and mean squared error (MSE) used as efficiency measures, illustrated in \reta{simus}. 

\begin{table}[ht]
	\centering
	\caption{MSE and bias of the BRI, ML and PWM estimators of the shape parameter, $\kappa$, from 5000 samples of different sizes.} \label{tab:simus}
	\begin{tabular}{c c lllllll}
		 & & \multicolumn{2}{c}{BRI} & \multicolumn{2}{c}{MLE}  & \multicolumn{2}{c}{PWM} \\
		$\kappa$ & $n$ & Bias & MSE & Bias & MSE & Bias & MSE \\ \toprule
		\multirow{3}{*}{1/3} & 15 & 0.135 & 0.865 & 0.475 & 1.365 & 0.691 & 16932 \\
		 & 50 & 0.036 & 0.191 & 0.129 & 0.218 & 11.75 & 6282 \\
		 & 100 & 0.018 & 0.095 & 0.063 & 0.102 & 1.096 & 1039 \\ \midrule
		 \multirow{3}{*}{3} & 15 & 0.014 & 0.011 & 0.052 & 0.016 & 0.802 & 13.88 \\
		 & 50 & 0.003 & 0.002 & 0.013 & 0.003 & 0.711 & 0.508 \\
		 & 100 & 0.002 & 0.001 & 0.007 & 0.001 & 0.688 & 0.474 \\ \midrule
		 \multirow{3}{*}{7} & 15 & 0.007 & 0.002 & -0.009  & 0.001 & -0.131 & 0.128 \\
		 & 50 & 0.002 & 0.0005 & -0.003 & 0.0004 & -0.039 & 0.030 \\
		 & 100 & 0.0009 & 0.0002 & -0.0015 & 0.0002 & -0.019 & 0.014
	\end{tabular}
\end{table}

Given that PWM works well only if (at least) the first two moments of the distribution exist it is not striking to confirm its poor performance for values of $\kappa \not \in (-0.2, 0.2)$ \citep{Hosking}, regardless of the sample size.  In contrast, both MLE and BRI estimators yield relatively low bias and MSE, even for moderate sample sizes, with their sampling distributions becoming increasingly similar as the sample size grows (\refi{simus}).  Both estimators are invariant under one-to-one transformations, while PWM is not; further, BRI credible intervals are invariant, a feature  we will exploit in the sequel.

\begin{figure}[ht]
	\centering
	\includegraphics[width=0.85\textwidth, keepaspectratio]{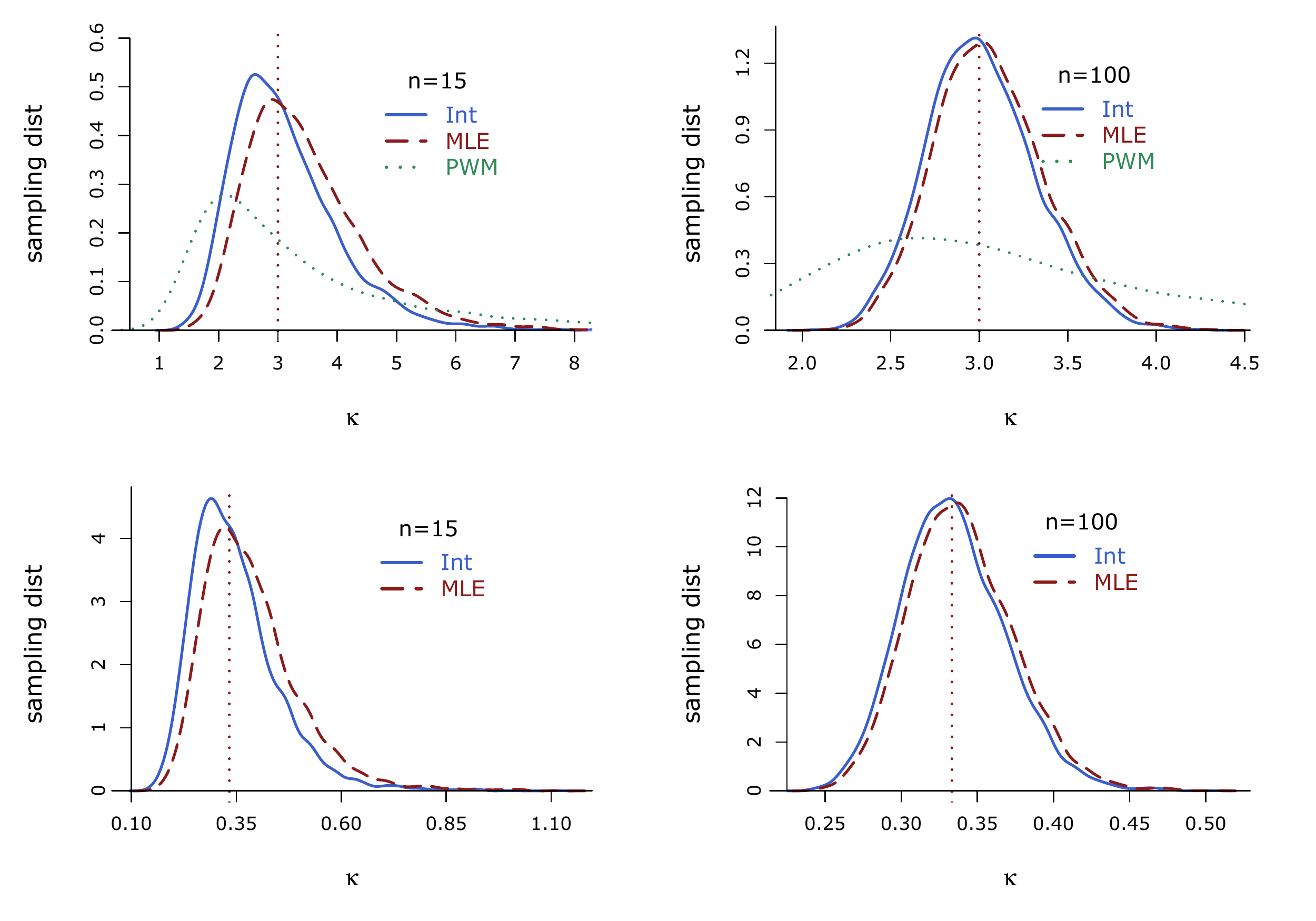}
	\caption{Sampling distributions of the BRI, ML and PWM estimators of the shape parameter, from 5000 simulations of sample sizes $n=15, 100$.  Top panels display the case $\kappa=3$ and the bottom when $\kappa=1/3$.  The latter do not display PWM due to its poor performance.  Note the one from the BRI estimator has a smaller variance than the other two alternatives for small sample size and is quite similar to the MLE when $n$ is large.} 	\label{fig:simus}
\end{figure}

\section{Bank equity capital requirements} \label{sec:real}

We apply the calibration approaches described above to equity risk capital that banks are required to hold in the banking Basel II regulations.  In the Basel II regulations in pghs.700 and 718 (LXXVI)\footnote{\url{http://www.bis.org/publ/bcbs128b.pdf}} a value-at-risk ($\vr$) approach is required for a 99th percentile one sided confidence interval on 10 day equity returns.  This means a bank will estimate what it thinks the 99th percentile 10 day equity returns can fall by (for example it might estimate this as a 10\% fall in its equities market value) and it is then required to hold at least this amount as a monetary capital amount on its balance sheet to demonstrate the bank can withstand a 99th percentile fall in the value of its equities.

The regulations mention a number of approaches are possible to calculate this VAR and in this case study a GPD is fitted to an historic time series of an equity index of 10 day returns.  The equity index used here is the FTSE 100 index taken from yahoo finance 2/4/1984-–26/7/2013\footnote{Available from \url{http://uk.finance.yahoo.com/q/hp?s=5EFTSE}}.

The raw data have been pre-processed to convert the daily index values into 10 day returns, $y_t$ (\ie the percentage change in value of the index over non overlapping 10 day periods).  A common question is either to use simple returns, $y_t = x_{t+10}/x_{t}-1$ where $x_{t}$ is the index value at time $t$, or log-returns, $y_t = \log\paren{x_{t+10}/x_{t}}$.   Whilst the difference between these two definitions is not crucial for a 10 day period, the log returns have been used in this case study.  This is because the left tail is of interest and is unbounded for the log return, but bounded at -100\% for the simple return.  An unbounded domain is potentially more appropriate for the GPD calibration.  When the 99th percentile has been calibrated in log returns, this needs to be converted back to a simple return for the VAR value.  For example if an -11\% fall in equity values is the log return 99th percentile, this is a $\exp(-0.11) - 1 = -0.104$ simple return fall in equity market value.

\subsection{Exploratory data analysis} \label{sub:subsection_name}

We explore some basic features of the data, presented for both simple and log returns on the left panel of \refi{fot}.
\begin{figure}[ht]
	\centering
	\includegraphics[width=0.875\textwidth, keepaspectratio]{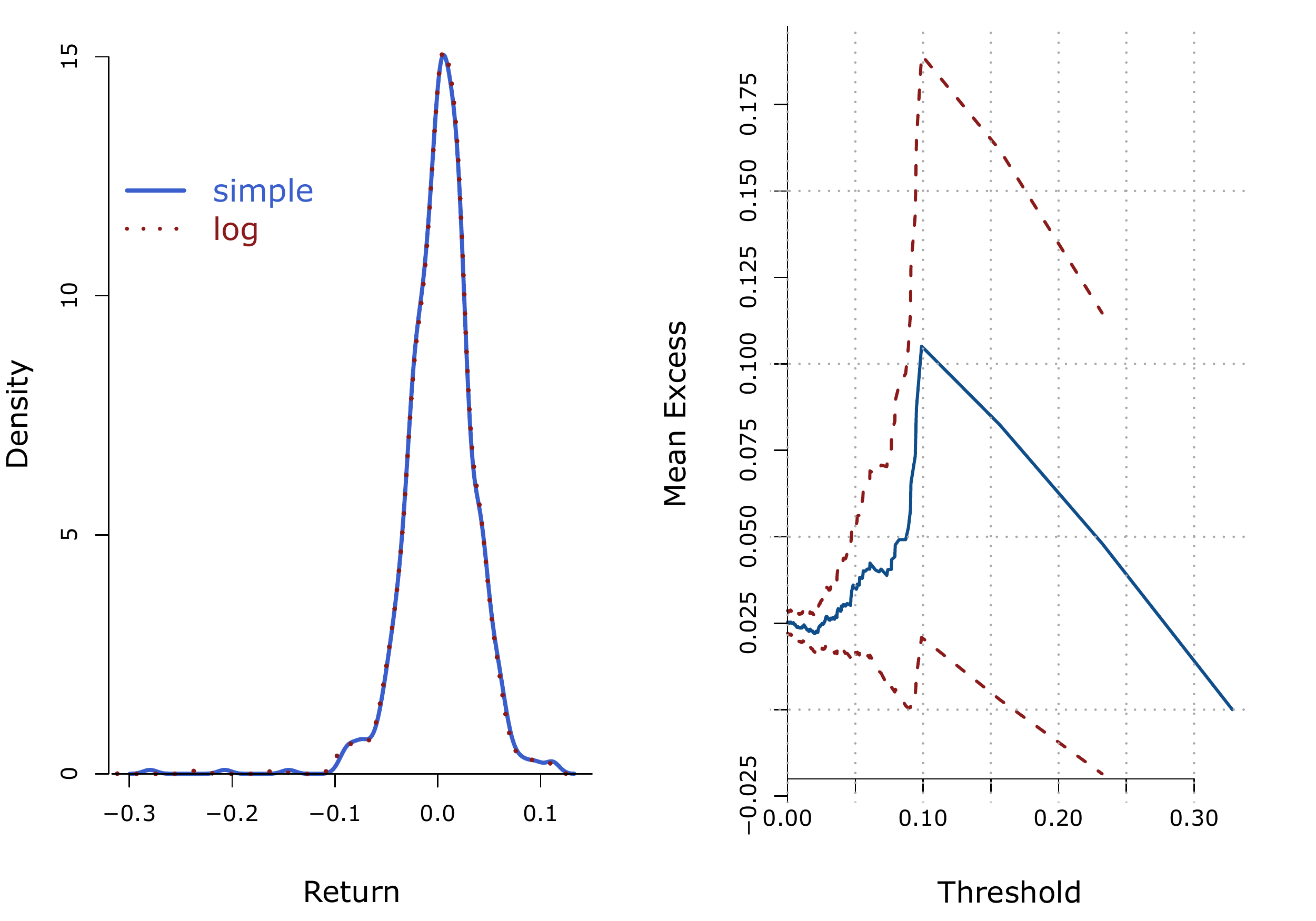}
	\caption{On the left panel, smoothed histograms of simple and log 10-day returns of the FTSE100 index from 2/4/84--26/7/13.  Both distributions are almost identical, left skew with heavy tails, and present a small secondary mode on the left hand tail.  On the right panel, Mean excess function of the negative 10-day returns of the FTS100 (solid) with 95\% confidence intervals (segmented).  The positive slope suggests a heavy tail, amenable to a power distribution.}
	\label{fig:fot}
\end{figure}

We would like to highlight that the distribution of the returns is fat tailed---it has a higher frequency of extreme events compared to a Gaussian distribution---as measured by its kurtosis (that of a Gaussian distribution is 3).  It is also negatively skewed---a higher proportion on events are on the left hand side of the mean, which emphasises the underlying financial risks.  Plotting the Mean Excess (ME) function, $M(u) = \ce{X-u}{x>u}$, is often used to explore whether the data has power tails \citep{Ghosh}.  A characteristic of a fat tailed GPD type distribution with negative shape parameter is an increasing straight line, while a decreasing line indicates thin tails; a horizontal line suggests exponential tails.  The right panel in \refi{fot} shows the mean excess plot of the absolute value of the negative log returns, which displays a positive slope (up to losses of about 10\%) suggesting a power distribution is appropriate for this data.  Combining this features with \refi{negrets} suggest a GPD with a negative shape parameter may be appropriate to model the negative returns of this data set.  
\begin{figure}[ht]
	\centering
	\includegraphics[width=0.75\textwidth, keepaspectratio]{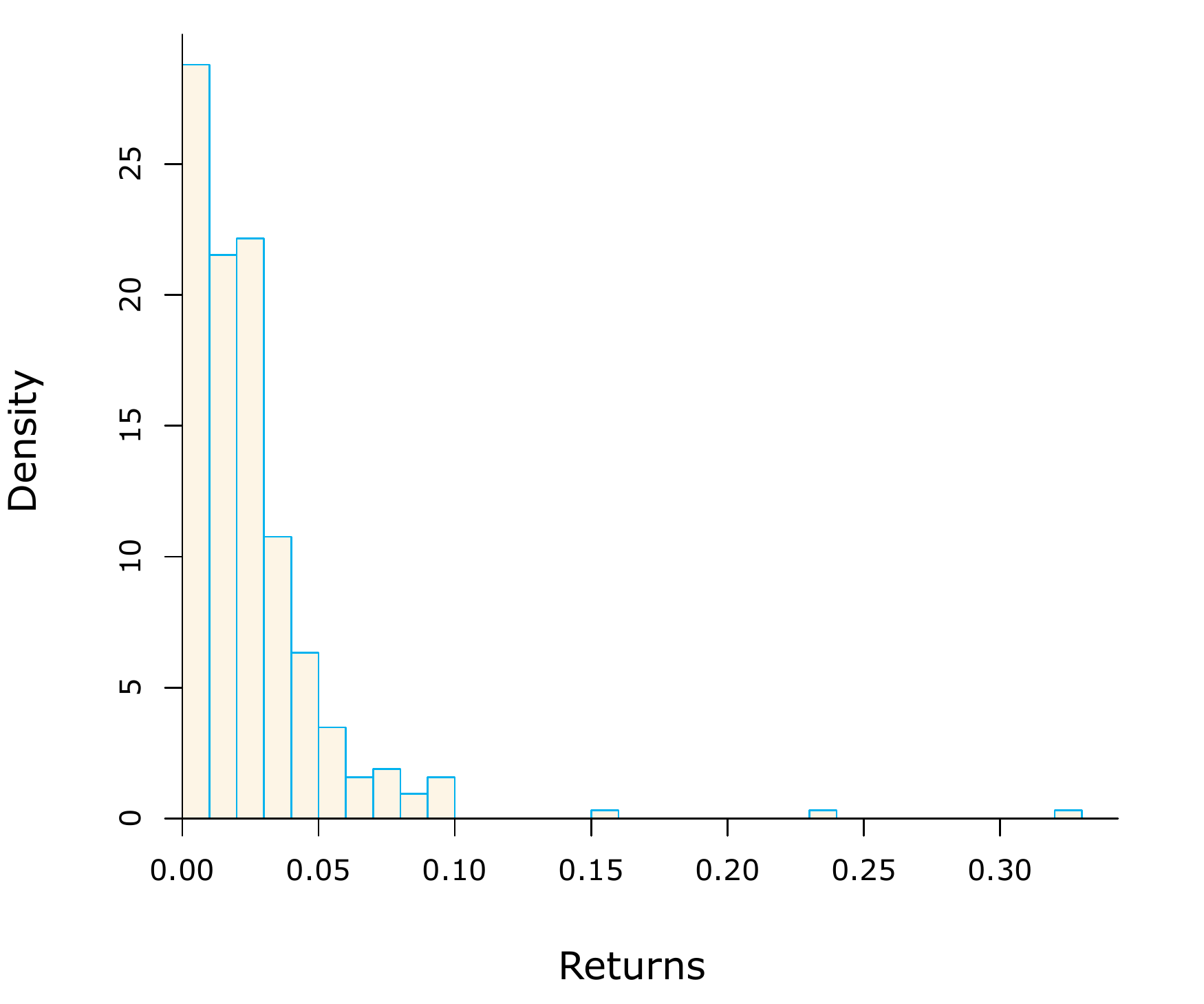}
	\caption{Histogram of the left tail of the 10-day log-returns from the FTSE100 data (in absolute value).  The shape shows a power decay and heavy tails, with some extreme events, suggesting a GPD may be a suitable  model.} 	\label{fig:negrets}
\end{figure}

\subsection{Peaks over threshold}\label{sub:peaks}

We now  apply the calibration approaches described in \rese{calib} and  \ref{sec:other} to the left tail of the 10-day FTSE100 log-returns,  using the peaks over threshold approach  \citep[][ with theory as in Section~\ref{ssub:pickands}]{frey}, which allows the focus to be on the percentiles of interest.

We do not discuss how to set the threshold, but refer the reader to \citet{zea2}.  \citet[][p.~280]{frey} suggest using the ME plot as a guide to threshold setting. We subjectively pick -5\% as the approximate point on the mean excess plot beyond which the slope appears to increase faster and end up with $n=33$.  It is noted there are still enough points beyond this level for reasonable calibration and the threshold is expected to be sufficiently high for the Picklands-Balkema-de Haan theorem to apply.

Relying on the fact that if $x \thicksim \gpd{-1/\alpha}{\beta/\alpha}$ then $y = 1/x \thicksim \ip[y]\alpha{1/\beta}$, we use the methods in \rese{calib} to obtain point and interval estimates of the shape parameter.  The MLE is straightforward to obtain, $\gorro \kappa = \log \paren{\gorro \beta/t}= 2.44$, where $t = \prod_{i=1}^n y_i^{1/n}$, is the geometric mean and $\gorro \beta = \max\samp ny = 19.71$ is the MLE of the shape.  The marginal posterior distribution of the shape parameter is \ga[\kappa]{n-1}{n/\gorro\kappa}, the BRI point estimate $\upd \kappa = 2.33$ and interval $(1.642, 3.298)$ are illustrated in \refi{jef}.  In particular, notice the intrinsic interval is different from the HPD, $(1.573, 3.195)$, highlight the fact that HPDs are not invariant under transformations, while the intrinsic is.  Given that the scale parameter typically is a nuisance parameter, we use the analytical approximation to the BRI estimator, $\upd \sigma \approx 2^{\frac{1}{n \gorro \kappa}} = 19.88.$

We use quasi-Newton \citep{Fletcher} to maximise the likelihood for the GPD, yielding MLEs $\gorro \kappa = -0.380$ and $\gorro \sigma =0.0222$.  The confidence intervals are found from the observed covariance matrix, which gives a standard errors for $\gorro\kappa$ of 0.228, thus an approximate 95\% confidence intervals for the shape is $(-0.828, 0.0679)$.  Using PWM, one gets $\tilde\kappa = -0.369$ and $\tilde \sigma = 0.0224$, with $(-0.844, 0.106)$ a CI of approximate 95\% for $\kappa$. Exploiting the invariance of the BRI estimator, $\upd \kappa = -0.429$ and $(-0.609, -0.303)$ the BRI interval of probability 0.95.   To fit the Bayesian model with Jeffreys prior, we generated chains of length $10^6$, dropped the first $10^4$ as burn-in and thinned every fifth draw, ending up with samples of size 198,000 for inference, the marginal posterior distributions are illustrated in \refi{jef}.  The posterior mean and median of the shape are $-0.254$ and $-0.252$, respectively and the equally tailed interval of posterior probability 0.95 is $(-0.472, -0.048)$.  Note both frequentist CIs include 0, suggesting an exponential tail behaviour, while the Bayesian alternatives strongly support heavy tails; also notice the intrinsic posterior has a smaller variance, hence the BRI interval is shorter than the equally tailed from Jeffreys prior.
\begin{figure}[ht]
	\centering
	\includegraphics[width=0.75\textwidth,keepaspectratio]{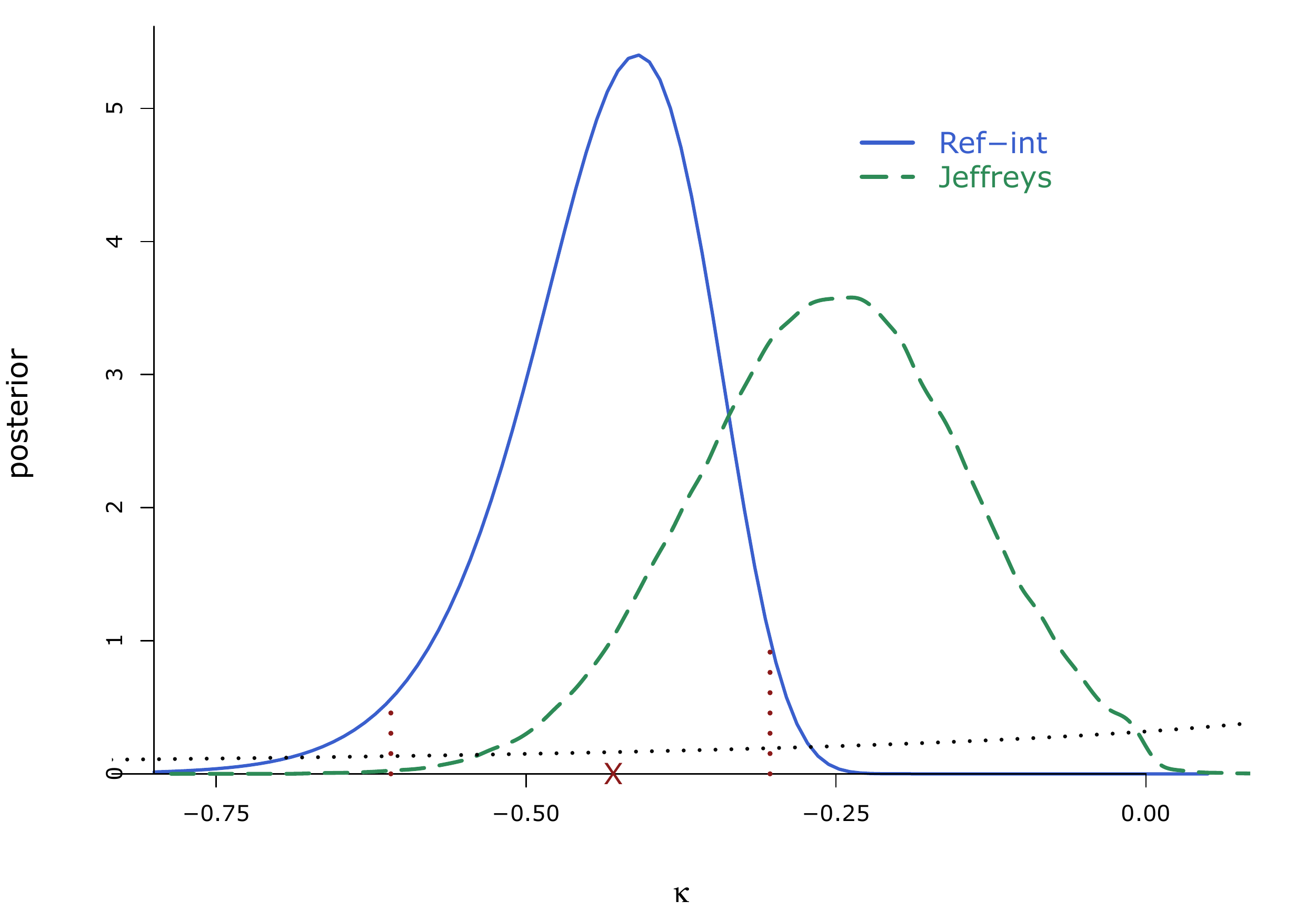}
	\caption{Marginal posterior distribution of the shape parameter from the FTSE100 data, thresholded at $-5\%$. Marked are the BRI estimate,  $\upd \kappa = -0.429$, and interval,  $(-0.609, -0.303)$ . The (proper) marginal Jeffreys prior is represented by the dotted line, the corresponding equally tailed interval of probability $0.95$ is $(-0.472, -0.051)$.}
	\label{fig:jef}
\end{figure}

The Pareto Principle and Gini index discussed in \rese{gini} depend only on the shape parameter, so we can use the invariance of the BRI and MLE approaches to calculate point and interval estimates shown in \reta{sumas}. As the calibration of the GPD is based on the tail of the data, we exploit its lack of memory to calculate
\[
	\vr_\varepsilon =   u + \frac \sigma\kappa \paren{1-\paren{\frac{1-\varepsilon}{\tilde F(u)}}^k }
\]
where $u= 0.05$ is the threshold and $\tilde F(u)$ is empirically estimated as the proportion of data points above the threshold relative to total number of data points \citep[][p.~283]{frey}; in our case $\tilde F(u)= 33/316$. 

\begin{table}[ht]
	\centering
	\caption{Point and interval estimates for the FTSE100 returns data, from the four approaches.  The middle point in the intervals is the corresponding point estimate.  Confidence intervals are of approximate 95\%, and Bayesian credible intervals of probability 0.95.  The point estimate from Jeffreys prior is the posterior median.} \label{tab:sumas}
	\begin{tabular}{lllr}
	 & \qquad \qquad $\kappa$  & \qquad Gini index & $\vr_{0.99} \%$\\ \toprule
	 BRI &   $(-0.609,-0.429, -0.303)$ & $(0.589, 0.637, 0.719)$ & $10.42$ \\
	 Jeffreys & $(-0.472, -0.253 , -0.051)$ & $(0.513, 0.573, 0.654)$ & $9.27$ \\
	 ML & $(-0.828, -0.380, 0.068)$ & $(0.484, 0.617, 0.853)$ & $9.99$ \\
	 PWM & $(-0.844, -0.403, 0.106)$ & $(0.475, 0.613, 0.865)$ & $9.04$
	\end{tabular}
\end{table}

It is worth noticing both frequentist approaches not rule out $\kappa=0$, while the Bayesian alternatives suggest $\kappa <0$, but all point estimates are negative.  The posterior distribution from the intrinsic approach is shifted to the right, compared to the Jeffreys alternative and has a smaller variance.  Frequentist confidence intervals are wider than the Bayesian credible counterparts.

\section{Final remarks} \label{sec:concs}

We have illustrated how the BRI approach can be used to calibrate the GPD by using a transformation from the inverted-Pareto distribution.  Four different approaches to calibrating the GPD have been presented.  Three of the approaches were compared in a simulation study of some simulated data from a Pareto distribution.  All four approaches were then compared for similarities and differences in a case study.

From the simulation study it is apparent that the repeated sampling behaviour of the PWM estimator is poor in general and some modification is needed if it is to work in practice \citep[see \eg][]{Chen}.  The results also indicate the BRI estimator has a lower MSE than the MLE even for moderated sample size, and also displays asymptotic Gaussianity (\refi{simus} and \reta{simus}).  Combined with its invariance under one-to-one transformations make it a competitive alternative for calibration.  One limitation of this simulation study was that it only simulated data from a Pareto distribution.  An extension might be to compare the calibration approaches for data simulated from other distributions.

The case study shows a practical example of how all four approaches can be used to calibrate the GPD.  Whilst there are some differences between the parameters for each of the four approaches, the point estimates of $\vr_{0.99}$ are roughly similar, suggesting a level between 9\% and 10\% for regulatory equity capital might appropriately meet the Basel II regulations.  Point estimates of the shape and scale (not shown) parameters are roughly similar for each calibration approach, barring Jeffreys, which shrinks the estimate towards the origin; however, the length of the interval Bayesian estimates are shorter than their frequentist counterparts. Moreover, frequentist interval estimates for $\vr_{0.99}$ are difficult to get, while those from Jeffreys prior are straightforward from the MCMC output.

The BRI approach has produced lower mean squared errors in the simulation studies, suggesting it to be more accurate for parameter estimation when the data is indeed from a Pareto distribution.  The MLE is relatively simple to understand and implement however there are questions over the convergence of the numerical optimisation methods, especially with fewer data points \citep{zea1}.  The PWM is likewise simple to understand and implement, but is efficient only for a subset of the parameter space and it could produce estimates with a likelihood of zero.

The MHA was left out of the simulation study as it is computationally intense.  The results in \resub{peaks} were obtained by running one million simulations.  It would not be possible to run to this level of accuracy and carry out an outer layer 1000 simulation analysis in a reasonable time period or without much greater computer power.  Also, implementing the sampler has a number of practical issues that are different for each data set, which may take time to resolve and ensure the MHA converges in a reasonable time period. However, our implementation is robust and may be used as an off-the-shelf option.

One area of interest that could be the subject of further work is how sensitive the results are to the threshold used for each calibration method.  A study might repeat the analysis looking at various different thresholds and how that impacts the shape, scale and $\vr_{0.99}$ for each threshold.

Another area of potential interest is how the time period of each data point impacts the shape parameter and GPD calibrations.  For example if the case study looked at equity returns over 1 day, 5 days, 1 year, etc. What would the impact be on the GPD calibration?  Clearly a large time step for each data point would be expected to give higher values for the 99th percentile, but would the Gini index remain invariant to other size time steps as for the 10 day period investigated in this case study?

\singlespace

\bibliographystyle{apalike}
\bibliography{SharpeJuarez_RefintGPD}

\begin{thebibliography}{}

\bibitem[Akhundjanov and Chamberlain, 2019]{land}
Akhundjanov, S.~B. and Chamberlain, L. (2019).
\newblock The power-law distribution of agricultural land size.
\newblock {\em Journal of Applied Statistics}, 0(0):1--13.

\bibitem[Behrens et~al., 2004]{Lopes}
Behrens, C.~N., Lopes, H.~F., and Gamerman, G. (2004).
\newblock Bayesian analysis of extreme events with threshold estimation.
\newblock {\em Statistical Modelling}, 4:227--244.

\bibitem[Beirlant et~al., 2005]{beirlant}
Beirlant, J., Goegebeur, Y., Teugels, J., and Segers, J. (2005).
\newblock {\em Statistics of Extremes: Theory and Applications}.
\newblock Springer, Chichester.

\bibitem[Berger et~al., 2009]{sun}
Berger, J., Bernardo, J.~M., and Sun, D. (2009).
\newblock The formal definition of reference priors.
\newblock {\em The Annals of Statistics}, 37:905–938.

\bibitem[Berger et~al., 2015]{sun2}
Berger, J., Bernardo, J.~M., and Sun, D. (2015).
\newblock Overall objective priors.
\newblock {\em Bayesian Analysis}, 10:189--221.

\bibitem[Bernardo, 2007]{intint}
Bernardo, J.~M. (2007).
\newblock Objective {Bayesian} point and region estimation in location-scale
  models.
\newblock {\em Statistics and Operations Research Transactions}, 31:3--44.

\bibitem[Bernardo and Juárez, 2003]{io}
Bernardo, J.~M. and Juárez, M.~A. (2003).
\newblock Intrinsic estimation.
\newblock In Bernardo, J.~M., Bayarri, M.~J., Berger, J.~O., Dawid, A.~P.,
  Heckerman, D., Smith, A. F.~M., and West, M., editors, {\em Bayesian
  Statitstics 7}, pages 465--476. University Press, Oxford.

\bibitem[Bernardo and Rueda, 2002]{rueda}
Bernardo, J.~M. and Rueda, R. (2002).
\newblock Bayesian hypothesis testing: a reference approach.
\newblock {\em International Statistical Review}, 70:351--372.

\bibitem[Castellanos and Cabras, 2007]{castellanos}
Castellanos, M.~E. and Cabras, S. (2007).
\newblock A default {B}ayesian procedure for the generalized {P}areto
  distribution.
\newblock {\em Journal of Statistical Planning and Inference}, 137:473--483.

\bibitem[Castillo and Hadi, 1997]{Castillo}
Castillo, E. and Hadi, A.~S. (1997).
\newblock Fitting the generalized {P}areto distribution to data.
\newblock {\em Journal of the American Statistical Association}, 92:1609--1620.

\bibitem[Castillo et~al., 2004]{Hadi}
Castillo, E., Hadi, A.~S., Balakrishnan, N., and Sarabia, J.~M. (2004).
\newblock {\em Extreme Value and Related Models with Applications in
  Engineering and Science}.
\newblock Wiley, New Jersey.

\bibitem[Chen et~al., 2017]{Chen}
Chen, H., Cheng, W., Zhao, J., and Zhao, X. (2017).
\newblock Parameter estimation for generalized {Pareto} distribution by
  generalized probability weighted moment-equations.
\newblock {\em Communications in Statistics - Simulation and Computation},
  46(10):7761--7776.

\bibitem[Coles, 2001]{Coles}
Coles, S. (2001).
\newblock {\em An Introduction to Statistical Modeling of Extreme Values}.
\newblock Springer-Verlag, London.

\bibitem[Coles et~al., 2003]{Pericchi}
Coles, S., Pericchi, L.~R., and Sisson, S. (2003).
\newblock A fully probabilistic approach to extreme rainfall modeling.
\newblock {\em Journal of Hydrology}, 273(1):35 -- 50.

\bibitem[Davidson and Smith, 1990]{davidson}
Davidson, A.~C. and Smith, R.~L. (1990).
\newblock Models for exceedances over high thresholds.
\newblock {\em Journal of the Royal Statistical Society B}, 52:393--442.

\bibitem[de~Haan and Ferreira, 2006]{Haan}
de~Haan, L. and Ferreira, A.~F. (2006).
\newblock {\em Extreme Value Theory. An Introduction}.
\newblock Springer, New York.

\bibitem[de~Zea~Bermudez and Kotz, 2010a]{zea1}
de~Zea~Bermudez, P. and Kotz, S. (2010a).
\newblock Parameter estimation of the generalized {P}areto distribution--{Part
  I}.
\newblock {\em Journal of Statistical Planning and Inference}, 140:1353--1373.

\bibitem[de~Zea~Bermudez and Kotz, 2010b]{zea2}
de~Zea~Bermudez, P. and Kotz, S. (2010b).
\newblock Parameter estimation of the generalized {P}areto distribution--{Part
  II}.
\newblock {\em Journal of Statistical Planning and Inference}, 140:1374--1388.

\bibitem[de~Zea~Bermudez et~al., 2009]{Mendes}
de~Zea~Bermudez, P., Mendes, J., Pereira, J. M.~C., Turkman, K.~F., and
  Vasconcelos, M. J.~P. (2009).
\newblock Spatial and temporal extremes of wildfire sizes in {Portugal}
  (1984–2004).
\newblock {\em International Journal of Wildland Fire}, 18:983--991.

\bibitem[de~Zea~Bermudez and Turkman, 2003]{Turkman}
de~Zea~Bermudez, P. and Turkman, M. A.~A. (2003).
\newblock {Bayesian} approach to parameter estimation of the generalized
  {Pareto} distribution.
\newblock {\em Test}, 12(1):259--277.

\bibitem[Diebolt et~al., 2005]{Diebolt2005}
Diebolt, J., El-Aroui, M.-A., Garrido, M., and Girard, S. (2005).
\newblock Quasi-conjugate {Bayes} estimates for {GPD} parameters and
  application to heavy tails modelling.
\newblock {\em Extremes}, 8(1):57--78.

\bibitem[Diebolt et~al., 2003]{Diebolt}
Diebolt, J., Guillou, A., and Worms, R. (2003).
\newblock Asymptotic behaviour of the probability weighted moments and
  penultimate approximation.
\newblock {\em ESAIM: Probability and Statistics}, 7:219--238.

\bibitem[Dupuis and Tsao, 1998]{Dupuis}
Dupuis, D. and Tsao, M. (1998).
\newblock A hybrid estimator for generalized {Pareto} and extreme-value
  distributions.
\newblock {\em Communications in Statistics - Theory and Methods},
  27(4):925--941.

\bibitem[Embrechts et~al., 1997]{Embrechts}
Embrechts, P., Klüppelberg, C., and Mikosch, T. (1997).
\newblock {\em Modelling Extremal Events}.
\newblock Springer-Verlag, Berlin.

\bibitem[Engeland et~al., 2004]{Engeland}
Engeland, K., Hisdal, H., and Frigessi, A. (2004).
\newblock Practical extreme value modelling of hydrological floods and
  droughts: A case study.
\newblock {\em Extremes}, 7(1):5--30.

\bibitem[Fawcett and Walshaw, 2006]{Fawcett}
Fawcett, L. and Walshaw, D. (2006).
\newblock A hierarchical model for extreme wind speeds.
\newblock {\em Journal of the Royal Statistical Society C}, 55:631--646.

\bibitem[Fletcher and Reeves, 1964]{Fletcher}
Fletcher, R. and Reeves, C.~M. (1964).
\newblock Function minimization by conjugate gradients.
\newblock {\em The Computer Journal}, 7(2):149--154.

\bibitem[Frigessi et~al., 2002]{Frigessi}
Frigessi, A., Haug, O., and Rue, H. (2002).
\newblock A dynamic mixture model for unsupervised tail estimation without
  threshold selection.
\newblock {\em Extremes}, 5(3):219--235.

\bibitem[Galambos, 1987]{gala2}
Galambos, J. (1987).
\newblock {\em The Asymptotic Theory of Extreme Order Statistics}.
\newblock Wiley, New York, 2 edition.

\bibitem[Ghosh and Resnick, 2010]{Ghosh}
Ghosh, S. and Resnick, S. (2010).
\newblock A discussion on mean excess plots.
\newblock {\em Stochastic Processes and their Applications}, 120:1492--1517.

\bibitem[Gilleland et~al., 2013]{Gilleland}
Gilleland, E., Ribatet, M., and Stephenson, A.~C. (2013).
\newblock A software review for extreme value analysis.
\newblock {\em Extremes}, 16:103--119.

\bibitem[Greenwood et~al., 1979]{Greenwood}
Greenwood, J.~A., Landwehr, J.~M., and Matalas, N.~C. (1979).
\newblock Probability weighted moments: Definition and relation to parameters
  of several distributions expressable in inverse form.
\newblock {\em Water Resources Research}, 15:1049--1054.

\bibitem[Holmes and Moriarty, 1999]{Holmes}
Holmes, J. and Moriarty, W. (1999).
\newblock Application of the generalized {Pareto} distribution to extreme value
  analysis in wind engineering.
\newblock {\em Journal of Wind Engineering and Industrial Aerodynamics},
  83(1):1 -- 10.

\bibitem[Hosking et~al., 1985]{Hosking}
Hosking, J. R.~M., Wallis, J.~R., and Wood, E.~F. (1985).
\newblock Estimation of the generalized extreme value distribution by the
  method of probability weighted moments.
\newblock {\em Technometrics}, 27:251--261.

\bibitem[Jagger and Elsner, 2006]{Jagger}
Jagger, T.~H. and Elsner, J.~B. (2006).
\newblock Climatology models for extreme hurricane winds near the {United
  States}.
\newblock {\em Journal of Climate}, 19(13):3220--3236.

\bibitem[Ju{\'a}rez and Schucany, 2004]{Schucany}
Ju{\'a}rez, S.~F. and Schucany, W.~R. (2004).
\newblock Robust and efficient estimation for the generalized {Pareto}
  distribution.
\newblock {\em Extremes}, 7(3):237--251.

\bibitem[Juárez, 2005]{Juarez}
Juárez, M.~A. (2005).
\newblock Objective {Bayes} estimation and hypothesis testing: the
  reference-intrinsic approach on non-regular models.
\newblock CRISM Working Paper 05-14.

\bibitem[Keylock, 2005]{Keylock}
Keylock, C. (2005).
\newblock An alternative form for the statistical distribution of extreme
  avalanche runout distances.
\newblock {\em Cold Regions Science and Technology}, 42(3):185 -- 193.

\bibitem[Koch, 2007]{Koch}
Koch, R. (2007).
\newblock {\em The 80/20 Principle: The Secret of Achieving More with Less}.
\newblock Nicholas Brealey Publishing, London.

\bibitem[Kotz and Nadarajah, 2000]{Kotz}
Kotz, S. and Nadarajah, S. (2000).
\newblock {\em Extreme Value Distributions: Theory and Applications}.
\newblock Imperial College Press, London.

\bibitem[Krehbiel and Adkins, 2008]{Krehbiel}
Krehbiel, T. and Adkins, L.~C. (2008).
\newblock Extreme daily changes in {U.S.} dollar {London} inter-bank offer
  rates.
\newblock {\em International Review of Economics and Finance}, 17(3):397 --
  411.

\bibitem[{La Cour}, 2004]{cour}
{La Cour}, B.~R. (2004).
\newblock Statistical characterization of active sonar reverberation using
  extreme value theory.
\newblock {\em IEEE Journal of Oceanic Engineering}, 29:310 -- 316.

\bibitem[Lana et~al., 2006]{Lana}
Lana, X., Burgueño, A., Martínez, M., and Serra, C. (2006).
\newblock Statistical distributions and sampling strategies for the analysis of
  extreme dry spells in {Catalonia (NE Spain)}.
\newblock {\em Journal of Hydrology}, 324(1):94 -- 114.

\bibitem[Leadbetter et~al., 1983]{leadbetter}
Leadbetter, M.~R., Lindgren, G., and Rootzen, H. (1983).
\newblock {\em Extremes and related properties of random sequences and
  processes}.
\newblock Springer, New York.

\bibitem[Lima et~al., 2016]{Lima}
Lima, C. H.~R., Lall, U., Troy, T., and Devineni, N. (2016).
\newblock A hierarchical {Bayesian} {GEV} model for improving local and
  regional flood quantile estimates.
\newblock {\em Journal of Hydrology}, 541:816 -- 823.

\bibitem[Malik, 1970]{malik}
Malik, H.~J. (1970).
\newblock Estimation of the parameters of a {P}areto distribution.
\newblock {\em Metrika}, 15:126--132.

\bibitem[McNeil, 1997]{McNeil}
McNeil, A.~J. (1997).
\newblock Estimating the tails of loss severity distributions using extreme
  value theory.
\newblock {\em ASTIN Bulletin}, 27(1):117–137.

\bibitem[McNeil et~al., 2005]{frey}
McNeil, A.~J., Frey, R., and Embrechts, P. (2005).
\newblock {\em Quantitative risk management: concepts, techniques and tools}.
\newblock University Press, Princeton.

\bibitem[Mendes et~al., 2010]{Mendes2010}
Mendes, J.~M., de~Zea~Bermudez, P.~C., Pereira, J., Turkman, K.~F., and
  Vasconcelos, M. J.~P. (2010).
\newblock Spatial extremes of wildfire sizes: {Bayesian} hierarchical models
  for extremes.
\newblock {\em Environmental and Ecological Statistics}, 17(1):1--28.

\bibitem[Moharram et~al., 1993]{Moharram}
Moharram, S.~H., Gosain, A.~K., and Kapoor, P.~N. (1993).
\newblock A comparative study for the estimators of the generalized {Pareto}
  distribution.
\newblock {\em Journal of Hydrology}, 150(1):169 -- 185.

\bibitem[Moisello, 2007]{Moisello}
Moisello, U. (2007).
\newblock On the use of partial probability weighted moments in the analysis of
  hydrological extremes.
\newblock {\em Hydrological Processes}, 21(10):1265--1279.

\bibitem[\"Oztekin, 2005]{Oztekin}
\"Oztekin, T. (2005).
\newblock Comparison of parameter estimation methods for the three-parameter
  generalized {Pareto} distribution.
\newblock {\em Turkish Journal of Agriculture and Forestry}, 29:419--428.

\bibitem[Pandey et~al., 2001]{Pandey2001}
Pandey, M.~D., Van~Gelder, P. H. A. J.~M., and Vrijling, J.~K. (2001).
\newblock The estimation of extreme quantiles of wind velocity using
  {L-moments} in the peaks-over-threshold approach.
\newblock {\em Structural Safety}, 23(2):179 -- 192.

\bibitem[Pandey et~al., 2004]{Pandey}
Pandey, M.~D., Van~Gelder, P. H. A. J.~M., and Vrijling, J.~K. (2004).
\newblock Dutch case studies of the estimation of extreme quantiles and
  associated uncertainty by bootstrap simulations.
\newblock {\em Environmetrics}, 15(7):687--699.

\bibitem[Peng and Welsh, 2001]{Peng}
Peng, L. and Welsh, A. (2001).
\newblock Robust estimation of the generalized {Pareto} distribution.
\newblock {\em Extremes}, 4(1):53--65.

\bibitem[Persky, 1992]{Persky}
Persky, J. (1992).
\newblock Retrospectives: {P}areto's law.
\newblock {\em The Journal of Economic Perspectives}, 6:181--192.

\bibitem[Pickands, 1975]{Pickands}
Pickands, J. (1975).
\newblock Statistical inference using extreme order statistics.
\newblock {\em The Annals of Statistics}, 3:119--131.

\bibitem[Pisarenko and Sornette, 2003]{Pisarenko}
Pisarenko, V.~F. and Sornette, D. (2003).
\newblock Characterization of the frequency of extreme earthquake events by the
  generalized {Pareto} distribution.
\newblock {\em Pure and Applied Geophysics}, 160(12):2343--2364.

\bibitem[Ragulina and Reitan, 2017]{Ragulina}
Ragulina, G. and Reitan, T. (2017).
\newblock Generalized extreme value shape parameter and its nature for extreme
  precipitation using long time series and the {Bayesian} approach.
\newblock {\em Hydrological Sciences Journal}, 62:863--879.

\bibitem[Robert, 1996]{Robert}
Robert, C.~P. (1996).
\newblock Intrinsic losses.
\newblock {\em Theory and Decision}, 40:191--214.

\bibitem[Rootzén and Tajvidi, 1997]{Rootzen}
Rootzén, H. and Tajvidi, N. (1997).
\newblock Extreme value statistics and wind storm losses: A case study.
\newblock {\em Scandinavian Actuarial Journal}, 1997(1):70--94.

\bibitem[Shi et~al., 1999]{Shi}
Shi, G., Atkinson, H., Sellars, C., and Anderson, C. (1999).
\newblock Application of the generalized {Pareto} distribution to the
  estimation of the size of the maximum inclusion in clean steels.
\newblock {\em Acta Materialia}, 47(5):1455 -- 1468.

\bibitem[Tancredi et~al., 2006]{Tancredi}
Tancredi, A., Anderson, C., and O'Hagan, A. (2006).
\newblock Accounting for threshold uncertainty in extreme value estimation.
\newblock {\em Extremes}, 9(2):87--106.

\bibitem[Zagorski and Wnek, 2007]{Zagorski}
Zagorski, M. and Wnek, M. (2007).
\newblock Analysis of the turbine steady-state data by means of generalized
  {Pareto} distribution.
\newblock {\em Mechanical Systems and Signal Processing}, 21(6):2546 -- 2559.

\end{thebibliography}
\begin{appendices}
\begin{sidewaystable}
	\centering
	\captionsetup{width=.9\linewidth} \caption{Some literature on GPD calibration approaches to data sets. Abbreviations are Maximum Literature (MLE), Method of Moments (MoM), Method of Medians (MM), Probability Weighted Moments (PWM), Elemental Percentile Method (EPM), Optimal Bias Robust Estimator (OBRE), Least Squares (LS), Maximum Entropy (ME), Minimum density power divergence estimator (MDPD)}\label{tab:paps}
	\small
	\begin{tabularx}{0.85\linewidth}{p{7cm} p{5cm} p{7cm}}
	\textbf{Reference} & \textbf{Area of application} & \textbf{Calibration types used} \\	\toprule
	\citet{Zagorski} & Engineering & MLE, LS to empirical mean excess function \\
	\citet{cour}  & Engineering & MLE \\
	\citet{Shi} & Engineering & MLE \\
	\citet{Hadi}  & Engineering & MOM, PWM, EPM \\
	\citet{Keylock} & Environment & MLE \\
	\citet{Krehbiel} & Economics & MLE \\
	\citet{Lopes} & Economics & Bayesian, ML \\
	\citet{Moisello} & Hydrology, Weather & PWM \\
	\citet{Oztekin}  & Hydrology & MOM, PWM. MLE, LS, ME \\
	\citet{Engeland} & Hydrology & PWM, MLE \\
	\citet{Pandey}  & Hydrology & L-Moments, MoM\\
	\citet{Castillo} & Hydrology & MOM, PWM, EPM \\
	\citet{Moharram} & Hydrology & MOM, PWM, MLE, LS \\
	\citet{castellanos} & Hydrology & Bayesian and MLE \\
	\citet{Tancredi} & Hydrology & Bayesian \\
	\citet{Diebolt} & Hydrology & Bayesian and  MLE \\
	\citet{Turkman} & Insurance, Hydrology & Bayesian, PWM and EPM \\
	\citet{Peng} & Hydrology & MM , MLE, OBRE \\
	\citet{Dupuis} & Hydrology & OBRE \\
	\citet{McNeil} & Insurance & MLE \\
	\citet{Rootzen} & Insurance & PWM, MLE \\
	\citet{Diebolt2005} & Insurance & Bayesian, MLE \\
	\citet{Frigessi} & Insurance & MLE \\
	\citet{Schucany} & Weather & MM, OBRE, MLE, MDPD \\
	\citet{Lana} & Weather & L-moments \\
	\citet{Jagger} & Weather & Bayesian and MLE \\
	\citet{Fawcett} & Weather & Bayesian and MLE \\
	\citet{Pericchi} & Weather & Bayesian and MLE \\
	\citet{Pandey2001} & Weather & L-moments, MoM\\
	\citet{Holmes} & Weather & LS to the empirical mean excess function \\
	\cite{Mendes} & Wildfires & PWM, MLE \\
	\citet{Mendes2010} & Wildfires & Bayesian \\
	\citet{Pisarenko} & Seismology & MLE
	\end{tabularx}
\end{sidewaystable}
\end{appendices}


\end{document}